\documentclass[10pt]{IEEEtran}

\usepackage{url}
\usepackage{graphics}
\usepackage{amsmath,amsfonts,amssymb,epsfig}
\usepackage{lscape}
\usepackage{latexsym}

\newcommand{\MF}{\mathbb{F}}

\newtheorem{theorem}{\bf Theorem}[section]
\newtheorem{lemma}[theorem]{\bf Lemma}
\newtheorem{proposition}[theorem]{\bf Proposition}

\newtheorem{definition}[theorem]{\sc Definition}
\newtheorem{conjecture}[theorem]{\sc Conjecture}

\begin{document}

\pagenumbering{gobble}

\title{On Cusick-Cheon's Conjecture About
Balanced Boolean Functions in the Cosets of the Binary Reed-Muller
Code}

\author{ Yuri L. Borissov
\thanks{Y. L. Borissov is with the Institute of Mathematics and Informatics,
 Bulgarian Academy of Sciences, Sofia 1113, Bulgaria.}
}

\maketitle

\begin{abstract}\noindent
It is proved an amplification of Cusick-Cheon's conjecture on
balanced Boolean functions in the cosets of the binary Reed-Muller
code $\it RM(k,m)$ of order $k$ and length $2^m$, in the cases
where $k = 1$ or $k \geq (m-1)/2$.
\end{abstract}


\section { Introduction}
For basic definitions and facts we refer to \cite{McWSl}. Let
${\it RM}(k,m)$ denote the $k$th-order Reed-Muller code of length
$2^m$. This linear code consists of all binary vectors of length
$2^m$ (truth tables) associated with Boolean functions in $m$
variables whose degree is less than or equal to $k$. A Boolean
function is said to be balanced if its truth table contains equal
number of zeroes and ones. We shall also call a truth table of a
balanced Boolean function balanced word. In \cite{CuChe}, the
authors have conjectured the following:
\begin{conjecture}\label{conj}
The code ${\it RM}(k,m)$, $k > 0$, considered as a coset in the
quotient space ${\it Q}(k,m) \stackrel{\rm def}{=} {\it
RM}(k+1,m)/{\it RM}(k,m)$ has more balanced functions than any
other coset in ${\it Q}(k,m)$.
\end{conjecture}
This conjecture was verified in cases $k =1, m-2$ \cite{Cu}. Based
on it the authors of \cite{CuChe} derived very good upper and
lower bounds on the number of balanced Boolean functions which are
contained in ${\it RM}(k,m)$. For some particular values of $k$
and arbitrary $m$, explicit formulas for the number of balanced
functions in ${\it RM}(k,m)$ are known \cite{McWSl}, \cite{Pe}.
Apart from trivial cases ($k=1,m-1$ and $m$) there are such
formulas for $k=2$ and $k = m-2$, as a part of the known
weight-distribution of the corresponding Reed-Muller codes.
However, the problem of determining the weight-distribution of
${\it RM}(k,m)$ in general, seems to be difficult \cite{McWSl} and
even partial results are welcomed \cite{KaToAz}. For similar
results in the context of cryptographic applications, see also
\cite{CCCS}, \cite{CaKl} and \cite[Ch. 8]{BP93}.

This paper is organized as follows. In next section we summarize
necessary background. In Section III we present  proofs of the
extension of Conjecture \ref{conj} for arbitrary coset of the
${\it RM}(k,m)$, if $k=1$ or $k \geq {(m-1)/2}$. Finally, in
Section IV we give an example which illustrates these
considerations.

\section{ Background}
Let us recall the so-called MacWilliams's identity \cite[p.
127]{McWSl}.
\begin{theorem}\label{McW}
Let ${\bf A}$ be a binary linear $(n,K)$ code and
$(A_{0},A_{1},\ldots,A_{n})$ denote the weight distribution of
${\bf A}$ i.e. the total number of vectors of  weight $i$ in ${\bf
A}$ is $A_{i}$ for each $i$. Then
\begin{equation}\label{eq1}
\sum_{i=0}^{n}A_{i}{\bf X}^{i} = 2^{K-n}
\sum_{i=0}^{n}B_{i}(1+{\bf X})^{n-i}(1-{\bf X})^{i},
\end{equation}
where $B_{i}$ is the total number of vectors of weight $i$ in
 $A^{\perp}$, the orthogonal code of ${\bf A}$.
\end{theorem}

We make use also of the following result proven by Assmus and
Mattson.
\begin{theorem}(\cite{AsMa})\label{ASMA}
 Let ${\bf A}$ be a binary linear $(n,K)$ code and ${\bf a}$ be an
 $n-$vector over ${\MF}_{2} = {\bf GF}(2)$ not in ${\bf A}$. Let
 $(d_{0},d_{1},\ldots,d_{n})$ denote the weight distribution of the
 coset ${\bf A}+{\bf a}$; thus the total number of vectors of
 weight $i$ in ${\bf A}+{\bf a}$ is $d_{i}$ for each $i$. Then
 \begin{equation}\label{eq2}
\sum_{i=0}^{n} d_{i}{\bf X}^{i} = 2^{K-n}
\sum_{i=0}^{n}(2b_{i}-B_{i})(1+{\bf X})^{n-i}(1-{\bf X})^{i},
 \end{equation}
where $b_{i}$ is defined as the number of vectors of weight $i$ in
the orthogonal code $A^{\perp}$ that are also orthogonal to ${\bf
 a}$ and $B_{i}$ is the total number of vectors of weight $i$ in
 $A^{\perp}$.
\end{theorem}
The above results were stated for a linear code over an arbitrary
finite field, but for our goals these particular versions are
enough.

The following deep theorem is due to McEliece.
\begin{theorem} (\cite{McE})
The weight of every codeword in ${\it RM}(k,m)$ is divisible by
$2^{[(m-1)/k]}$.
\end{theorem}

Let us remind also the following definition.
\begin{definition}\cite[p. 151]{McWSl}
For an arbitrary positive integer $n$ the Krawtchouk polynomial
$P_{k}({\bf x};n) = P_{k}({\bf x})$ is defined as
\begin{eqnarray*}
P_{k}({\bf x};n) \stackrel{\rm def}{=} \sum_{j=0}^{k}(-1)^{j}
{{\bf x} \choose j} {n-{\bf x} \choose k-j},
\end{eqnarray*}
$k=0,1,2,\ldots$, where as usual ${\bf x}$ is a variable while the
binomial coefficients are defined as in Ex. 18 \cite[Ch. 1]
{McWSl}.
\end{definition}
Note that $P_{k}({\bf x})$ is a polynomial of degree $k$.

For the sake of completeness we recall  the definitions of weight
and Walsh-Hadamard transform of a Boolean function. Below,
"$\sum$" stands for the ordinary summation, while "$+$" is used
for the modulo-$2$ summation.

The {\it weight} of a Boolean function $f$ is equal to the number
of nonzero positions in the truth table of $f$ and is denoted by
$wt(f)$. A Boolean function $f$ is uniquely determined by its
Walsh-Hadamard transform, which is a real-valued function over
${\MF}_{2}^{m}$ defined for all $\bf {\omega}\in {\MF}_{2}^{\rm
m}$ as
\begin{eqnarray}\label{Wal}
W_f({\bf \omega}) = \sum_{\bf {x} \in {\MF}_{\rm 2}^{\rm m}}
(-1)^{f({\bf x}) + {\bf x}\cdot {\bf \omega}}=2^{m}-2wt(f+{\bf
x}\cdot {\bf \omega})\,,
\end{eqnarray}
Here the {\it dot product} or scalar product of the vectors ${\bf
x} = (x_1, x_2,\ldots, x_m)$ and ${\bf \omega}=(\omega_1,
\omega_2, \ldots, \omega_m)$ is defined as ${{\bf x} \cdot {\bf
\omega}} = x_1 \omega_1 + x_2 \omega_2 +  \cdots +  x_m \omega_m$.

It is easy to see that the Boolean function $f$ is balanced if and
only if $W_f({\bf 0}) = 0$. We recall also, the so-called
Parseval's equation:
\begin{equation}
\sum_{\bf {\omega} \in {\MF}_{2}^{\rm m}}W_f({\bf \omega})^{2} =
2^{2m}
\end{equation}

\section{ The Proofs}

First, we shall prove the following lemma.
\begin{lemma}\label{Krawtchouk}
For an arbitrary even positive integer $n$ and $i=0,1,\ldots,n$
let us define the numbers $K(i,n)$ as
\begin{eqnarray*}
K(i,n) \stackrel{\rm def}{=}\sum_{j=0}^{i}(-1)^{j} {i \choose j}
{n-i \choose n/2-j},
\end{eqnarray*}
Then $K(i,n)$ is equal to $0$ for $i$ odd, negative when $i \equiv
2\; ({\rm mod}\; 4)$ and positive when $i \equiv 0\; ({\rm mod}\;
4)$.
\end{lemma}

\begin{proof}
Note that $K(i,n)$ is actually $P_{n/2}(i)$, where $P_{n/2}({\bf
x})$ is the Krawtchouk polynomial of degree $n/2$. Further we make
use of the Ex. $46$ \cite[p. 153]{McWSl} which states that for
arbitrary nonnegative integers $i$ and $j$ the following recurrent
formula holds:
\begin{eqnarray*}
(n-i) P_{j}(i+1) = (n-2j)P_{j}(i) - i P_{j}(i-1),
\end{eqnarray*}
where  $P_{j}({\bf x})$ is the Krawtchouk polynomial of degree
$j$. In our case $j = n/2$, thus we have:
\begin{eqnarray}\label{eq3}
(n-i) P_{n/2}(i+1) = - i P_{n/2}(i-1)
\end{eqnarray}
The initial values: $P_{n/2}(0) = {n \choose n/2}$ and $P_{n/2}(1)
= 0$ are easily computed (see e.g. equation $5.57$ and Ex. $44$
\cite[pp. 151-153]{McWSl}). The proof follows by induction on $i$
using recurrent relation $(\ref{eq3})$.
\end{proof}

Now, we shall prove an amplification of Cusick-Cheon's conjecture
in some special cases.
\begin{theorem}\label{basic}
Let $B(k,m)$ be the number of balanced words in the binary
Reed-Muller code ${\it RM}(k,m), k \geq {(m-1)/2}$. Then any
nontrivial coset of ${\it RM}(k,m)$ contains less than $B(k,m)$
balanced words.
\end{theorem}

\begin{proof}
Let ${\bf a}$ be a binary vector of length $n = 2^{m}$ not in
${\bf A} = {\it RM}(k,m)$  and ${\bf C} = {\bf A} + {\bf a}$ be
the considered coset. It is well-known that the dimension of ${\bf
A}$ is $K = \sum_{j=0}^k {m \choose j}$ and the orthogonal code
${\bf A}^{\perp}$ coincides to ${\it RM}(m-k-1,m)$. Let $b_{i}$ be
the number of vectors of weight $i$ in ${\bf A}^{\perp}$ that are
orthogonal to ${\bf a}$ and $B_{i}$ is the total number of vectors
of weight $i$ in ${\bf A}^{\perp}$, $0 \leq i \leq n$. Applying
{Theorem} \ref{McW} and {Theorem} \ref{ASMA}, we get,
respectively:
\begin{equation*}
 B(k, m) = 2^{K - n} \sum_{i=0}^n B_i K(i, n)
\end{equation*}
\begin{equation*}
 d_{n/2} = 2^{K - n} \sum_{i=0}^n (2b_i - B_i) K(i, n),
\end{equation*}
where $d_{n/2}$ is the number of balanced words in ${\bf C}$ and
the numbers $K(i,n)$ are defined in Lemma \ref{Krawtchouk}. So, we
yield:
\begin{equation}\label{eq6}
B(k,m) - d_{n/2} = 2^{K - n+1} \sum_{i=0}^n (B_i - b_i) K(i, n)
\end{equation}
Clearly, by definition of the numbers $B_{i}$ and $b_{i}$, we
have: $B_{i} \geq b_{i}$. Also, there exists at least one weight
$i$ for which last inequality holds strictly, since, otherwise the
vector ${\bf a}$ must belong to ${\bf A}$. Furthermore, if $ k
\geq {(m-1)/2}$ then $(m-1)/(m - k - 1) \geq 2$, and hence
according to McEliece's Theorem all weights of codewords in ${\bf
A}^{\perp}$ are divisible by 4. Thus, by Lemma \ref{Krawtchouk}
the numbers $K(i,n)$ are positive and consequently, the sum in
equation (\ref{eq6}) is positive as well, which completes the
proof.
\end{proof}

Finally, we shall prove the following extension of the Conjecture
\ref{conj} in the case where $k=1$:
\begin{proposition}
Any nontrivial coset of the first order binary Reed-Muller code
${\it RM}(1,m)$ contains less than $2^{m+1}-2$ balanced words.
\end{proposition}
\begin{proof}
First, let us note that the number of balanced words in ${\it
RM}(1,m)$ itself, is $2^{m+1}-2$, and the two unbalanced words are
the all-zero and all-one vectors of length $2^{m}$.

Let $f$ be an arbitrary non-affine function and ${\bf f}$ be its
corresponding truth table. We consider the coset ${\bf C} = {\it
RM}(1,m)+{\bf f}$. By the Parseval's equation there exists at
least one ${\bf \omega}$, say ${\bf \omega}_0$, such that
$W_f({\bf \omega}_0) \not = 0$. Let $g = f+{\bf x}\cdot {\bf
\omega}_0$. Clearly, $W_f({\bf \omega}_0) = W_g({\bf 0})$ and
therefore the function $g$ is unbalanced, as well as $g+1$, of
course. Suppose, $g$ and $g+1$ are the only two unbalanced
functions (words) in ${\bf C}$. Then obviously, $W_f({\bf \omega})
= 0$, for ${\bf \omega} \not = {\bf \omega}_0$ and by the
Parseval's equation $W_f({\bf \omega}_0) = \pm 2^{m}$. Hence,
according to (\ref{Wal}), $wt(g)$ is equal to either $0$ or
$2^{m}$, which means that either $f = {\bf x}\cdot {\bf \omega}_0$
or $f = {\bf x}\cdot {\bf \omega}_0+ 1$, a contradiction to a
choice of $f$. Consequently, ${\bf C}$ contains more than two
unbalanced words which completes the proof.
\end{proof}
\section{An Example}

In this section we present an example which illustrates the above
considerations. We shall use the same notations as in the proof of
{Theorem} \ref{basic}.

Consider the $(m-2)$th order Reed-Muller code ${\it RM}(m-2,m), m
\geq 3,$ which is in fact the extended Hamming code of length
$n=2^{m}$. The orthogonal code is the first-order Reed-Muller code
${\it RM}(1,m)$ and consists of truth tables of the affine
functions and the vectors ${\bf 0}$,${\bf 1}$. So, the nonzero
$B$'s are: $B_{2^{m-1}} = 2^{m+1}-2$ and $B_{i} = 1$ for
$i=0,2^{m}$. Applying equation (\ref{eq1}) for the
weight-distribution of the code ${\bf H}_{m} = {\it RM}(m-2,m)$,
we get the well-known (see e.g. \cite{Pe}):
\begin{eqnarray*}
\sum_{i=0}^{n} H_{i}{\bf X}^{i} = 2^{-(m+1)}[(1+{\bf
X})^{n}+(2^{m+1}-2)(1-{\bf X}^{2})^{n/2} +\\ (1-{\bf X})^{n}]
\end{eqnarray*}
Thus, we have:
\begin{equation*}
B(m-2,m) = H_{n/2} = {1 \over n}[{n \choose n/2} + (n-1){n/2
\choose n/4}]
\end{equation*}
Let ${\bf a}_{1}$ be the following $2^{m}-$vector of weight $2$: $
(0,0,\ldots,1,1)$. This vector is associated with the Boolean
function which is a product of the first $m-1$ amongst the Boolean
variables $Y_{1},Y_{2}, \ldots ,Y_{m-1},Y_{m}$ i.e. the function:
$Y_{1}Y_{2} \ldots Y_{m-1}$. It is easy to see that the truth
table of an affine function is orthogonal to ${\bf a}_{1}$ only if
this function does not contain $Y_{m}$ as an essential variable.
So, $b_{2^{m-1}} = 2^{m}-2$ and since the vectors ${\bf 0}$, ${\bf
1}$ are orthogonal to ${\bf a}_{1}$ it follows $b_{0} = b_{2^{m}}
= 1$. Applying equation (\ref{eq2}) for the weight-distribution of
the coset ${\bf C}_{1} = {\bf H}_{m}+{\bf a}_{1}$, we get:
\begin{eqnarray*}
\sum_{i=0}^{n} d_{i}{\bf X}^{i} = 2^{-(m+1)}[(1+{\bf
X})^{n}-2(1-{\bf X}^{2})^{n/2} +(1-{\bf X})^{n}]
\end{eqnarray*}
Therefore, we have:
\begin{equation*}
d_{n/2} = {1 \over n}[{n \choose n/2}-{n/2 \choose n/4}]
\end{equation*}
This result coincides with the outcome of computations given in
\cite{Cu}. In fact, it can be shown that all cosets of ${\it
RM}(m-2,m)$ in ${\it RM}(m-1,m)$ are affine equivalent (see, e.g.
\cite{Ho}) and therefore they have the same weight-distribution
(in particular, the same number of balanced functions).

Let, now ${\bf a}_{2}$ be the following $2^{m}-$vector of weight
$1$: $(0,0,\ldots,0,1)$. This vector is associated with the
Boolean function $Y_{1}Y_{2} \ldots Y_{m-1}Y_{m}$. Of course, we
can proceed as in the previous case, but the following simple
arguments show that every word in the coset ${\bf C}_{2} = {\bf
H}_{m}+{\bf a}_{2}$ is with odd weight. Indeed, if ${\bf f} \in
{\bf H}_{m}$ then $wt({\bf f})$ is an even number and $wt({\bf
f}+{\bf a}_{2})$ is equal to $wt({\bf f}) \pm 1$ accordingly to
the value of the last coordinate of ${\bf f}$. Thus, there are no
balanced functions in the coset ${\bf C}_{2}$, and by similar
arguments this is valid also for all cosets of ${\it RM}(m-2,m)$
not in ${\it RM}(m-1,m)$.

\section{ Conclusion}
In this paper, we consider an extension of Cusick-Cheon's
conjecture on balanced Boolean functions in the cosets of the
binary Reed-Muller code ${\it RM}(k,m)$ and prove it in the
special cases: $k = 1$ or $k \geq {(m-1)/2}$. To our knowledge,
the Conjecture \ref{conj} is still unproved (or disproved) in the
remaining cases. Note also, that Theorem \ref{basic} is valid for
any code of even length whose orthogonal is doubly-even code i.e.
if the weights of all codewords in the orthogonal code are
divisible by 4.
\section*{\it Acknowledgments}

The author wishes to thank Thomas W. Cusick for pointing out the
problem and useful discussions.


\begin{thebibliography}{99}
\bibitem{AsMa}
E.F. Assmus Jr. and H.F. Mattson Jr., "The Weight-Distribution of
a Coset of a Linear Code", {\it IEEE Transactions on Information
Theory}, 1978, pp. 497.
\bibitem{CCCS}
P. Camion, C. Carlet, P. Charpin, and N. Sendrier, "On
Correlation-Immune Functions", {\it Crypto} 1991, LNCS vol. 576,
Springe-Verlag, pp. 86-100, 1992.
\bibitem{CaKl}
C. Carlet and A. Klapper, "Upper Bounds on the Numbers of
Resilient Functions and Bent Functions", 23rd Symposium on
Information Theory in the Benelux, Louvain-La-Neuve, Belgique, may
2002.
\bibitem{CuChe}
T.W. Cusick and Y. Cheon, "Counting Balanced Boolean Functions in
$n$ Variables with Bounded Degree", {\it Experimental Mathematics
}, 16:1, pp. 101-105.
\bibitem{Cu}
T.W.Cusick's talk at NATO Advanced Study Institute "Boolean
Functions in Cryptology and Information Security", Moscow,
September 8-15, 2007.
\bibitem{Ho}
X.D. Hou, "$GL(m,2)$ Acting on $R(r,m)/R(r-1,m)$", {\it Discrete
Math.}, vol. 149, pp. 99-122, 1996.
\bibitem{KaToAz}
T. Kasami, N. Tokura, and S. Azumi, "On the Weight Enumeration of
Weights Less than 2.5d of Reed-Muller Codes", {\it Information and
Control}, vol. 30, pp. 380-395, 1976.
\bibitem{McWSl}
F.J. McWilliams and N.J.A. Sloane, {\it The Theory of
Error-Correcting Codes}, North-Holland Publishing Company 1977.
\bibitem{McE}
R.J. McEliece,"Weight Congruences for $p-$ary Cyclic Codes", {\it
Discrete Math.}, 3 (1972), pp. 177-192.
\bibitem{Pe}
W.W. Petersen, {\it Error-Correcting Codes}, John Wiley and Sons
Inc., 1961.
\bibitem{BP93}
B.~Preneel, {\it Analysis and Design of Cryptographic Hash
Functions}, Ph. D. thesis, Katholieke Universiteit Leuven, 1993.
\end{thebibliography}
\end{document}